\documentclass[journal]{IEEEtran}
\UseRawInputEncoding
\pdfoutput=1
\usepackage{hyperref}
\usepackage{bm}
\usepackage{mathrsfs}
\usepackage{amsmath}
\usepackage{amsthm}

\usepackage{paralist}
\usepackage{multirow}
\usepackage{graphicx}
\usepackage{amsfonts}
\usepackage{amsmath,epsfig}
\usepackage{mathtools}
\usepackage{stfloats}
\usepackage{amssymb}
\usepackage{epstopdf}
\usepackage{latexsym}
\usepackage{color}
\usepackage{cite}
\usepackage{algorithm}
\usepackage{algorithmic}

\ifCLASSINFOpdf

\else

\fi

\begin{document}
%

\title{Task Offloading with Multi-Tier Computing Resources in Next Generation Wireless Networks}%: Enabling Technologies and Future Challenges}

\author{Kunlun Wang, \IEEEmembership{Member,~IEEE}, Jiong Jin, \IEEEmembership{Member,~IEEE}, Yang Yang, \IEEEmembership{Fellow,~IEEE}, Tao Zhang, \IEEEmembership{Fellow,~IEEE}, Arumugam Nallanathan, \IEEEmembership{Fellow,~IEEE}, Chintha Tellambura, \IEEEmembership{Fellow,~IEEE}, and Bijan Jabbari \IEEEmembership{Fellow,~IEEE}
%                 Jun Li, \IEEEmembership{Senior~Member,~IEEE}, Yang Yang, \IEEEmembership{Fellow,~IEEE}, Wen Chen, \IEEEmembership{Senior~Member,~IEEE}, and Lajos Hanzo,~\IEEEmembership{Fellow,~IEEE}

%\thanks{The work of K. Wang was supported by the National Natural Science Foundation of China (NSFC) under grant 61801463. This work of J. Li was supported by National Key R\&D Program under Grant 2018YFB1004800.
%The work of Y. Yang was supported in part by the National Key Research and Development Program of China under grant 2019YFB1803304, and the National Development and Reform Commission of China (NDRC) under grant "5G Network Enabled Intelligent Medicine and Emergency Rescue System for Giant Cities".
%The work of W. Chen was supported by the NSFC under grant 61671294.
%L. Hanzo would like to acknowledge the financial support of the Engineering and Physical Sciences Research Council projects EP/N004558/1, EP/P034284/1, EP/P034284/1, EP/P003990/1 (COALESCE), of the Royal Society's Global Challenges Research Fund Grant as well as of the European Research Council's Advanced Fellow Grant QuantCom.}

\thanks{K. Wang is with the Shanghai Key Laboratory of Multidimensional Information Processing, East China Normal University, Shanghai 200241, China, and also with the School of Communication and Electronic Engineering, East China Normal University, Shanghai 200241, China. (e-mail: klwang@cee.ecnu.edu.cn).}
\thanks{J. Jin is with the School of Science, Computing and Engineering Technologies, Swinburne University of Technology, Melbourne, Australia. (email: jiongjin@swin.edu.au).}
\thanks{Y. Yang is with Terminus Group, Beijing 100027, China, ShanghaiTech University, Shanghai 201210, China, and Peng Cheng Laboratory, Shenzhen 518055, China. (e-mail: dr.yangyang@terminusgroup.com).}
\thanks{T. Zhang is with the US National Institute of Standards and Technology (NIST). (e-mail: taozhang1@yahoo.com).}
\thanks{A. Nallanathan is with the School of Electronic Engineering and Computer Science at Queen Mary University of London, UK. (email: a.nallanathan@qmul.ac.uk).}
\thanks{C. Tellambura is with the Department of Electrical and Computer Engineering, University of Alberta, Canada. (e-mail: ct4@ualberta.ca).}
\thanks{B. Jabbari is with the Department of Electrical and Computer Engineering, George Mason University, Fairfax, VA, USA. (e-mail: bjabbari@gmu.edu).}
}
\maketitle

        % <-this % stops a space
%\thanks{The authors are with Department of Electronic Engineering, Shanghai Jiao Tong University, and SEU National Key Laboratory on Mobile Communications,
%China; (e-mail:\{kunlun1228\emph{};wenchen\}@sjtu.edu.cn)}
%\thanks{X.~Wang is with DOCOMO Beijing Communications Laboratories Co., Ltd~(e-mail:wangxl@docomolabs-beijing.com.cn).}

%\thanks{This work is supported by NSF China \#60972031, by national 973 project \#2012CB316106 and
%\#2009CB824904, by national huge special project \#2012ZX03004004,
%by national key laboratory project \#ISN11-01, by Huawei Funding
%\#YBWL2010KJ013, and by Foundation of GuangXi
%University~\#XGL090033.}

%\markboth{IEEE Wireless Communications Letters ,~Vol.~1, No.~1, January~2012}%

%{Shell \MakeLowercase{\textit{et al.}}: Bare Demo of IEEEtran.cls for Journals}

\begin{abstract}
With the development of next-generation wireless networks, the Internet of Things (IoT) is evolving towards the intelligent IoT (iIoT), where intelligent applications usually have stringent delay and jitter requirements. In order to provide low-latency services to heterogeneous users in the emerging iIoT, multi-tier computing was proposed by effectively combining edge computing and fog computing. More specifically, multi-tier computing systems compensate for cloud computing through task offloading and dispersing computing tasks to multi-tier nodes along the continuum from the cloud to things.
In this paper, we investigate key techniques and directions for wireless communications and resource allocation approaches to enable task offloading in multi-tier computing systems. A multi-tier computing model, with its main functionality and optimization methods, is presented in details. We hope that this paper will serve as a valuable reference and guide to the theoretical, algorithmic, and systematic opportunities of multi-tier computing towards next-generation wireless networks.

\end{abstract}
\IEEEpeerreviewmaketitle
%============================================================================================================================
\section{Introduction}

%As Fifth Generation (5G) mobile networks are being rolled out, the telecom industry and academia are now coordinating the 6G research effort towards defining the requirements and use cases for Beyond 5G (B5G) or so-called Sixth Generation (6G) mobile networks. 6G will be more encompassing in terms of communication requirements in contrast to its predecessor, being more society centric in terms of requirements; in addition to the vertical market requirement, it is widely accepted that the 6G drive will be influenced by global policy on sustainability goals for an ageing and growing population, as well as addressing societal challenges. The aim is to deliver a 6G architecture that promotes digital inclusion and accessibility, as well as unlocking economic value and opportunities in rural communities.

%Despite the ongoing rollout of 5G, a key bottleneck to ultra-high speed is the underlying spectrum. However, driven the market demand towards higher bit rates to entertain virtual reality (VR) applications requires higher frequencies over the THz band (0.1¨C10 THz) that will be key to ubiquitous 6G networks. In particular, THz frequencies have the potential to deliver ample spectrum, over hundred gigabits-persecond (Gbps) data rates, massive connectivity, denser networks, and highly secure transmissions.

As the fifth generation wireless networks (5G) being commercially deployed, research efforts of the sixth generation wireless networks (6G) have begun to define 6G requirements and use cases. Four promising use cases have emerged.
First, holographic telepresence allows realistic, full motion, three-dimensional (3D) images of people and objects to be projected as holograms into a meeting room to interact with each other in real time \cite{2018-David-6G,2022-Ali-6G}. Such remote holographic meeting, surgery, or distant learning will reduce the need for travel.
The second key use case is digital twin, which creates a real-time, comprehensive, and detailed digital (virtual) copy of a physical object, or system \cite{2019-Zong-6G}. %Once the twin passed all the tests, the physical model can be replaced realtime.
Digital twins help push the boundaries of system reliability, used to support a wide range of capabilities such as diagnostics and fault prediction. The third one is connected industrial robots, tactile Internet and intelligent cars. In this use case, the components of a control system (e.g., controllers, sensors, and actuators) are distributed across a wide geographic region \cite{2019-Zong-6G}, and therefore need to be connected via a wide area mobile infrastructure. In addition, these intelligent applications usually require stringent delay and jitter performance, with typical maximum tolerable network latency below $1$ milliseconds. The fourth use case is automated network operation empowered by distributed artificial intelligence (AI), intelligent IoT (iIoT), and big data technologies~\cite{2020-Yang-RIS,2017-Zhang-vehicular}. %Therefore, there are a wide range of new computation-intensive and time-critical applications as augmented reality~(AR), \textcolor{black}{smart manufacturing}, smart agriculture and intelligent robotic cars, etc,~\cite{2020-Yang-RIS,2017-Zhang-vehicular} of B5G for third parties to run.

\begin{figure*}[!t]
\begin{center}
\includegraphics [width=5.8in]{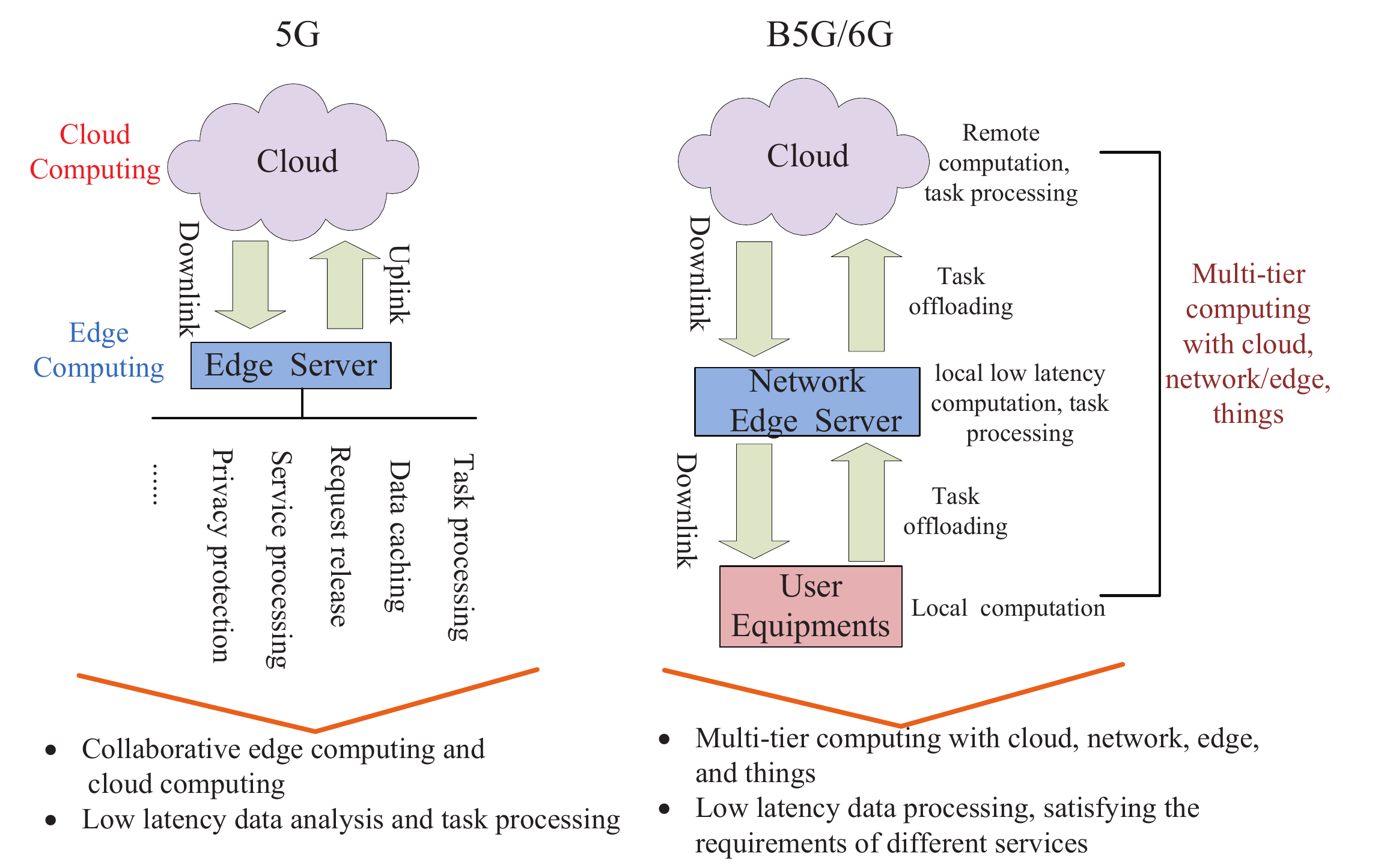}
\caption{Edge computing versus multi-tier computing} \label{MCIoT}
\end{center}
\end{figure*}

Many current and future applications require %With the iterative development of next-generation wireless communications, AI, big data and cloud computing, the ultimate development goal of the Internet of Things (IoT) will move from the internet of everything to the intelligent internet of everything, or iIoT~\cite{2019-Sergey-IIOT}.
low latency, high reliability, and high data security protection~\cite{2019-Sergey-IIOT}. These cannot be adequately met by the traditional cloud computing model, which requires to upload massive data and computing tasks to the cloud through fronthaul links and hence is difficult to meet the requirements of low latency and high energy efficiency. To provide low-latency services, a new computing paradigm called multi-tier computing was proposed by effectively combining edge computing and fog computing \cite{2016-Chiang-Fog,2018-Yang-MEETS}. With multi-tier computing, a large number of smart devices with varying computational resources, located around the end user, can communicate and cooperate with each other to execute computational tasks. A comparison between multi-tier computing and the current 5G-based edge computing is illustrated in Fig.~\ref{MCIoT}. Multi-tier computing complement cloud computing and edge computing by offloading and dispersing computational (and communication and caching) tasks and resources along the continuum from the cloud to things.

%

%According to the model of multi-tier computing systems and vision of next generation wireless networks \cite{2022-Letaief-edge},
Multi-tier computing is expected to become an important part of 5G systems and beyond \cite{2019-Yang-Nature,2022-Letaief-edge}, supporting computational-intensive applications that require low latency but high energy efficiency, high reliability and high security services. The effectiveness of multi-tier computing depends largely on resource scheduling among edge and cloud nodes to reduce service latency and ease network congestion \cite{2018-Yang-MEETS,2019-Wang-TVT,2020-Wang-MMIMO,2021-Wang-IRS}. Because of the above characteristics, multi-tier computing plays a vital role in many industrial applications of iIoT, such as connected cars, smart grids, smart buildings, and smart cities etc. Along the development of next-generation wireless networks, all kinds of user equipments (UEs) will be online all the time, promoting the advancement of iIoT and bringing diversified applications. These novel intelligent applications typically require low latency and demand prompt computations for real-time task processing and high data rates.
However, mobile devices often have limited computation, storage, and energy resources. To overcome these limitations, the capability of offloading computational tasks from the end users to nodes in the multi-tier computing systems will be essential. Such %, as an emerging technology, has been proposed for sticking a compelling compromise between the resource-constrained nature of compact devices and their high-complexity tasks.
task offloading enables distributed smart devices to share their idle computation and storage resources, which realizes the efficient utilization of multi-dimensional resources for low latency task processing.
Additionally, multi-tier computing systems will provide new task offloading models with the development of B5G wireless communications system, even 6G wireless communication system, and the new generation of embedded AI. As computational capability moves from the cloud to edge and UEs, the computing capabilities will be integrated into the network. Computing and network will be deeply integrated. Cloud-to-things computing capabilities should also be coordinated, leading to a new stage of intelligent multi-tier computing systems.
%However, a large scale of terminals and intelligent services pose many challenges to multi-tier computing networks.

%---------------------------------------------------------------------------------------------------------------------------
\subsection{Task Offloading in Multi-Tier Computing-based Next-Generation Wireless Networks}
\begin{figure*}[!t]
\begin{center}
\includegraphics [width=5.8in]{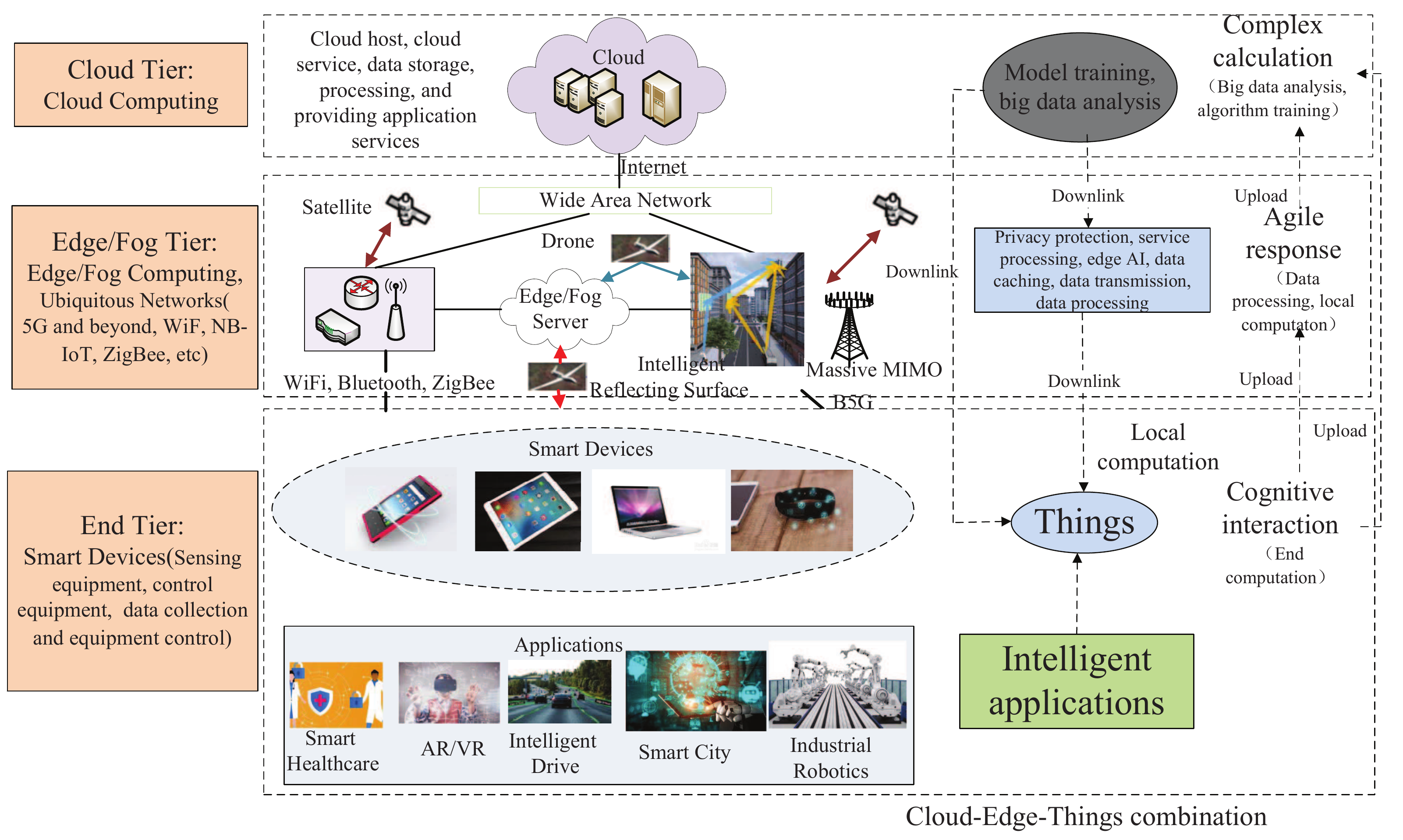}
\caption{Illustration of multi-tier computing network.} \label{Fig-system_model}
\end{center}
\end{figure*}

%The idea of caching popular content at the edge of the network is gaining momentum as one of the most promising enablers of next-generation networks~\cite{2014-Boccardi-5G,2015-Li-caching,2017-Avik-fog}.
Next-generation wireless communication networks present various novel technologies, including massive multiple-input multiple-output (MIMO), intelligent reflecting surface (IRS), space-air-ground integrated networks (SAGIN) and edge AI, etc. A multi-tier computing model integrates these radio technologies and AI to reduce task execution latency, allows large-scale user access, and enables efficient task offloading to realize efficient collaborative computing and multi-dimensional communication, caching, computation resource coordination. An example of multi-tier computing-based next-generation networks is illustrated in Fig.~\ref{Fig-system_model}. Basically, it consisting of two types of nodes, i.e., task node (TN) and helper node (HN). In particular, multiple TNs are able to offload their tasks to multiple HNs. It remains a fundamental challenge to effectively map multiple tasks or TNs into multiple HNs to minimize the total cost, such as task offloading latency or energy consumption, in a distributed manner, known as the multi-task multi-helper (MTMH) problem~\cite{2019-Yang-POMT,2020-Liu-POST}.
%In all, these core technologies of the next generation wireless networks will improve the performance of task offloading in multi-tier computing systems.

Massive MIMO can provide array gains, diversity gains, and multiplexing gains without increasing spectrum and power resources. It has been shown in \cite{2014-Lu-mMIMO} that massive MIMO schemes improve significantly the data rates at the cell edge and also increase exponentially the spectrum efficiency, resulting in an order of magnitude increasing of system capacity. The integration of multi-tier computing and massive MIMO has been proven to enhance task offloading performance in terms of ultra reliability and low latency~\cite{2019-Ozgun-FogMIMO,2019-Hessam-FogMIMO,2018-Chen-FogMIMO,2018-Ratheesh-FogMIMO}.
%As expected, the integration of fog computing and massive MIMO can enhance the performance of task offloading in multi-user fog computing systems~\cite{2019-Ozgun-FogMIMO,2019-Hessam-FogMIMO,2018-Chen-FogMIMO,2018-Ratheesh-FogMIMO}.
In particular, Bursalioglu \MakeLowercase{\textit{et al.}}~\cite{2019-Ozgun-FogMIMO} proposed and analyzed an architecture, so called fog massive MIMO, where a large number of
multi-antenna base stations (BSs) is densely deployed and serves the users using zero-forcing beamforming~(ZFBF). %In~\cite{2019-Ozgun-FogMIMO}, Bursalioglu \MakeLowercase{\textit{et al.}} proposed and analyzed an architecture named "fog massive MIMO", in which a large number of multi-antenna BSs are densely deployed, and zero-forcing beamforming~(ZFBF) is used to serve the users.
Pirzadeh \MakeLowercase{\textit{et al.}} \cite{2019-Hessam-FogMIMO} investigated the viability of supervised learning for user location estimation based on network signals transmitted by the users. In~\cite{2018-Chen-FogMIMO}, Chen investigated a specific edge computing mechanism for fronthaul-constrained distributed Massive MIMO systems, aiming to minimize energy consumption on user devices. In \cite{2018-Ratheesh-FogMIMO}, Mungara \MakeLowercase{\textit{et al.}} considered a new architecture termed as dense fog massive MIMO, where the users establish high-throughput and low-latency data links in a seamless and opportunistic manner, as they travel through a dense fog with large number of multiple antennas remote radio heads (RRHs). Although the above research demonstrates the advantages of massive MIMO-based multi-tier computing, the influence of channel estimation error on resource allocation and task offloading is not studied, which is particularly important for time-varying multi-tier computing systems.
%On the other hand, as the fog systems provide additional computing capabilities at the edge of the network, a major question that they raise is how to manage task execution. More precisely, how to decide which tasks to be executed in the end-users stratum, the fog stratum, and the cloud stratum. On a finer level, the dilemma is which nodes a particular task should be assigned to.
On the other hand, to compensate for cloud computing, multi-tier computing systems provide computational capabilities both at the edge and center of the network. However, one of the major issues is how to manage task offloading and execution. More specifically, how to decide which tasks to perform at the end-user, fog/edge, or in the cloud. At a more granular level, the issue boils down to which node a particular task should be assigned to.

%A variety of sophisticated wireless communication technologies have been proposed for next generation wireless networks, including massive MIMO and intelligent reflecting surface (IRS).
In B5G, the radio frequency may exceed $6$ Gigahertz. Since higher-frequency signal is more sensitive to the blockage by obstacles, the coverage of each base station will be significantly reduced \cite{2017-Lien-5G,2015-Bai-Mwave}. Furthermore, devices at the cell edge or behind obstacles suffer from low task offloading rates, increasing both delay and energy consumption of task offloading in multi-tier computing systems~\cite{2021-Bai-IRSWP}.
IRS, regarded as an effective auxiliary wireless network technology for potentially achieving high spectrum and energy efficiency via low-cost reflecting elements, has attracted increasing attention to circumvent these restrictions and is listed as one of the candidate key technologies in 6G by academia and industry \cite{2019-Wu-IRS,2020-Di-IRS,2020-Marco-RIS,2020-Wu-IRST,2021-Bai-IRSMEC,2021-Hu-IRS,2021-Wu-IRSNOMA}.
Due to the combination of array aperture gain (achieved by combining a direct transmission signal with an IRS reflection signal) and the reflection-assisted beamforming gain (achieved by controlling the phase shift of each IRS element), IRS is able to improve the success rate of task offloading and the potential of efficient resource scheduling of multi-tier computing systems. %For array aperture gain, it is usually achieved by combining a direct transmission signal with an IRS reflection signal. And the beamforming gain is achieved by controlling the phase shift of each IRS element, for enhancing the offloading rate of the devices at the cell edge.
Therefore, IRS will be a key technology for task offloading in next-generation wireless networks. \textcolor{black}{In \cite{2021-Chu-IRSEd,2021-Wang-IRS}, the authors have studied the impact of IRS on computational performance in a multi-tier computing system, which have demonstrated the benefits of the IRS to improve the task offloading, in comparison to the benchmark schemes.}

Meanwhile, by integrating satellite systems, aviation systems and ground communication systems, SAGIN is widely treated as a cornerstone of future 6G network. This new architecture supports seamless and near-instantaneous hyper-connectivity \cite{2021-Dong-STNMEC}, aiming at global data acquisition with high temporal and spatial resolution, high-precision real-time navigation and positioning, and broadband wireless communications. Being an essential component of SAGIN, UAVs are deployed flexibly at the air-network layer, capable of assisting terrestrial network in task offloading and communications/computing/caching resources management due to their flexibility and proximity~\cite{2018-Cheng-AGIMEC}. However, even with efficient task offloading, it is still not trivial to meet the quality of experience (QoE) requirements of heterogeneous users in the SAGIN.

Because of increasingly complex wireless networks, a typical 5G node is expected to have $2000$ or more configurable parameters. Therefore, a recent new trend is to optimize task offloading and wireless resource allocation through AI technologies~\cite{2017-Mao-MEC,2018-Mao-DL}, including applying AI at multiple protocol layers (e.g., physical layer resource allocation, data link layer resource allocation, and traffic control) \cite{2018-Tang-DL}.
%In view of the increasing complexity of mobile networks, e.g., a typical 5G node is expected to have $2000$ or more configurable parameters, a recent new trend is to optimize wireless communication by Artificial Intelligence (AI) techniques \cite{2017-Mao-MEC,2018-Mao-DL}, include but not limited to the application of AI for Physical Layer (PHY), Data Link Layer, and traffic control \cite{2018-Tang-DL}.
Thanks to the rapid development of mobile chipsets, the computational capabilities of edge devices have been substantially improved. For example, smart devices nowadays have as much computational capability as computing servers a decade ago. In addition, edge servers could provide end users with low latency AI services that are not possible to achieve directly on the devices. Since the computational resources of edge servers are not as much as those of cloud centers, it is necessary to adopt joint design principles across edge servers and edge devices to reduce task execution latency and enhance privacy for task offloading \cite{2017-Mao-MEC}. As a result, advances in multi-tier computing systems offer an opportunity to move the frontiers of AI from the cloud center to the edge of the network, inspiring a new field of research called edge AI, including both AI model training and inference procedures.

Distributed AI and federated learning algorithms are performed on multi-tier computing servers at the access network, which is capable of realizing low-latency task processing and providing computing, storage, and networking services. Since the data is processed at the edge server in the close proximity of smart device by task offloading, there is no need to transfer a large amount of raw data to the back-end server. Thus, using edge AI on task offloading infrastructure not only saves network bandwidth on backhaul links, but also reduces greatly the task execution latency. %In fact, the edge network will migrate to the end user's local area, including those devices in proximity.
Edge AI will be a significant step towards reducing task execution latency by intelligently enabling task offloading and local caching of popular file and content migration. In addition, intelligent task offloading for computational tasks will make it possible to further virtualize users' handsets and improve battery lifetime.
%In concurrent development, federated learning allows multiple parties to jointly train a deep learning model on their combined data without any participants having to reveal their data to a centralized server.
In all, edge AI provides a new paradigm of optimization algorithms design for efficient task offloading and service-driven resource allocation in multi-tier computing systems \cite{2022-Letaief-edge}. By seamlessly integrating sensing, communications, computing and intelligence, edge AI will empower multi-tier computing systems to support multiple intelligent applications, including industrial robots, intelligent robotic cars, and intelligent healthcare etc.

\subsection{Main Contributions}

In this paper, a vision of multi-tier computing with intelligent task offloading is presented. Furthermore, we summarize its interactions with various wireless techniques and resource allocations, as well as discuss future research directions and open problems, to embrace the era of multi-tier computing based next-generation wireless networks.

Against the above backdrop, our contributions could be further detailed as follows:
\begin{itemize}
\item \textcolor{black}{The vision, challenges and solutions for task offloading in multi-tier computing systems towards next-generation wireless networks.} %In order to maximize the EE of content delivery, we optimize the probability distribution of the cached files.
\item \textcolor{black}{The task offloading in multi-tier computing systems is presented, including the massive MIMO-aided task offloading, the task offloading with IRS, the task offloading in Space-Air-Ground Integrated Networks (SAGIN), and edge AI-empowered task offloading.}
\item The resource allocation for task offloading is elaborated. Specifically, we introduce the main functionality and optimization methods as well as the algorithms for task offloading in multi-tier computing systems.
\item  We discuss the research directions and open problems of task offloading for multi-tier computing-based next-generation wireless networks.
    %, and the prior works in this area did not explicitly take into account the effect of battery life in the energy efficiency.
\end{itemize}

\subsection{Paper Organization}
The rest of the paper is organized as follows. Section \uppercase\expandafter{\romannumeral2} introduces the enablement of multi-tier computing for next-generation wireless networks, while Section \uppercase\expandafter{\romannumeral3} presents the resource allocation for multi-tier computing systems. Section \uppercase\expandafter{\romannumeral4} is focused on research directions and open problems for multi-tier computing. In Section \uppercase\expandafter{\romannumeral5}, we provide our conclusions. %Table \ref{table} lists the frequently used notations.

\section{Enablement of Task Offloading for Multi-Tier Computing-based Next-Generation Networks}
In this section, we present the vision, challenges and solutions for task offloading in multi-tier computing systems, including the massive MIMO-aided task offloading, the task offloading with IRS, the task offloading
in SAGIN, and edge AI-empowered task offloading.
\subsection{Massive MIMO-Aided Task Offloading}
%With the advent of next generation wireless standards, new high performance technologies have been introduced. One of these key technologies is constituted of massive Multiple-input multiple-output~(MIMO) systems \cite{2014-Andrews-5G}, which are being increasingly adopted in different networking and computing frameworks.
With the advent of next-generation of wireless standards, new high-performance technologies are introduced. One of these key technologies is massive MIMO \cite{2014-Andrews-5G} that has been increasingly adopted in different networking and computing frameworks. However, the works of~\cite{2016-Pu-D2Dfog,2017-Chen-D2D,2018-Yang-MEETS,2018-Yang-DEBTS} mainly considered single-antenna computation offloading systems, by taking joint resources allocation and task offloading into account, but failed to exploit the MIMO advantages in terms of task offloading efficiency.
As we know that MIMO techniques have the potential of achieving high spectral efficiency (SE)~\cite{2002-Paulraj-MIMO,2016-Wang-MUMIMO,2015-Wkl-packet}, so as to improve the task offloading data rate. Therefore, new technologies have been introduced to improve the performance of edge users from current levels. In particular, equipping the base stations~(BSs) with a large number of antennas, widely known as massive MIMO, has emerged as one of the most promising solutions \cite{2010-Marzetta-MIMO,2013-Fredrik-MIMO} to significantly improve systems SE and energy efficiency trade-off.
%More specifically, when the number of antennas increases, the channels become more deterministic, which is referred to as channel hardening. Hence, the achievable data rates are mostly determined by large-scale fading, and so is the resource allocation. This means that there is no need to frequently update the resource allocation, yielding substantial savings in the signalling overhead. In all, massive MIMO schemes increase the spectral and energy efficiencies and support an increased number of users, both of which are crucial for fog computing systems.
More specifically, as the number of antennas increases, channels become more deterministic, known as channel hardening. Data rates and communication resource allocations are hence largely determined by large-scale fading. This implies that resource allocation does not need to be updated frequently, resulting in significant savings in signal transmission overhead. In summary, massive MIMO schemes improve spectrum and energy efficiency and support an increased number of users, both of which are critical for multi-tier computing systems.

In addition, as the core technology of wireless communication, relay technique has been integrated into various wireless communication standards to improve network coverage and throughput \cite{2009-Liu-Relay}. In particular, massive MIMO-enabled relay networks can enhance spectral efficiency and achieve more reliable data transmission for spatially distributed user nodes through intermediate massive antenna relay nodes \cite{2015-Amarasuriya-MIMO,2012-Yang-MIMO}.
Thus, a massive MIMO-aided fog access node~(FAN) serving as a relay is capable of significantly improving the data rate of offloaded tasks and the task execution efficiency. The new computing model that combines massive MIMO with multi-tier computing will facilitate efficient task offloading of computation-intensive tasks to achieve efficient collaborative computing and multi-dimensional communication, caching, computation resource scheduling.

\begin{figure}[!t]
\begin{center}
\includegraphics [width=3.6in]{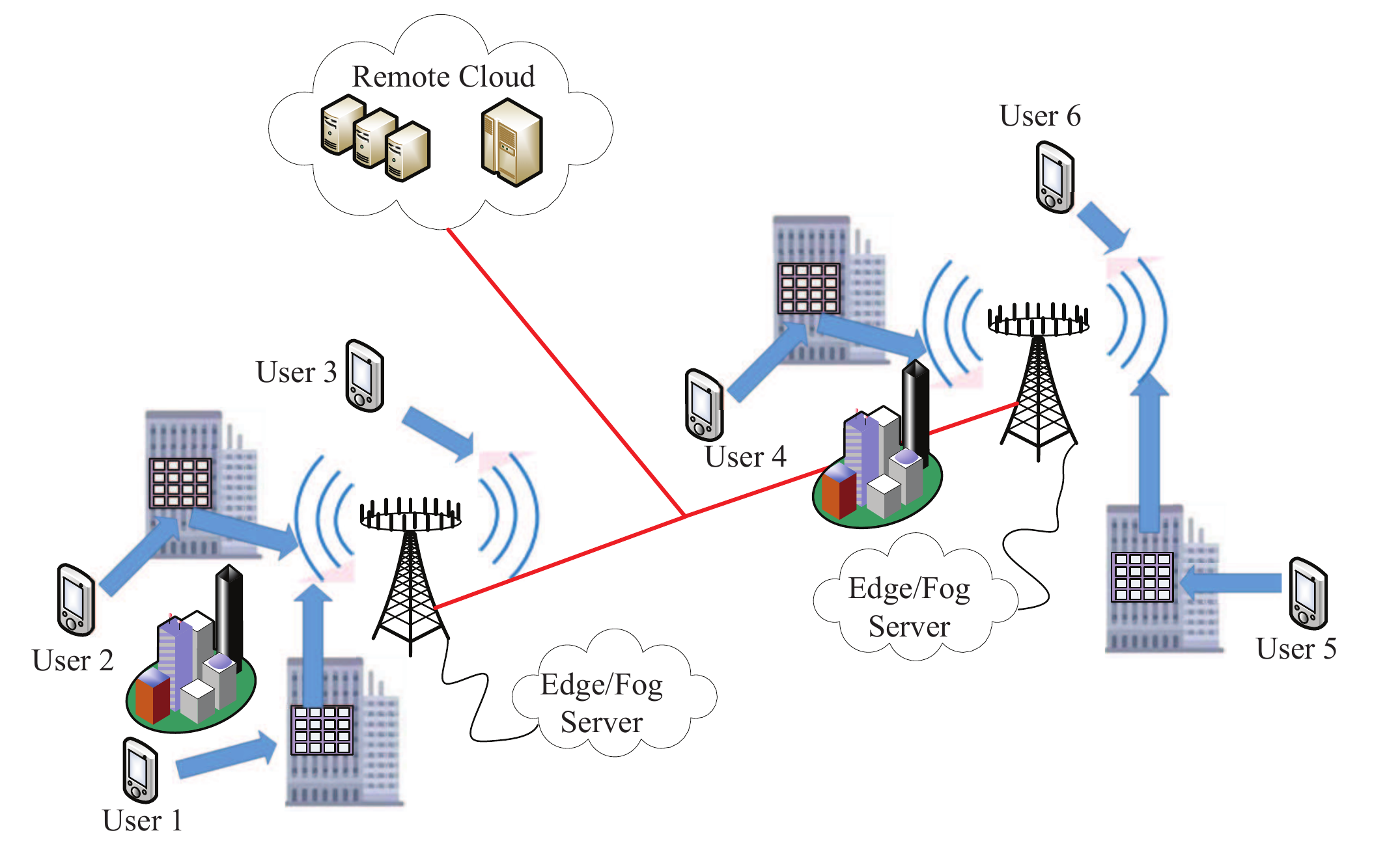}
\caption{Illustration of a massive MIMO and IRS-enabled multi-tier computing network.} \label{MMI}
\end{center}
\end{figure}

\subsection{Task Offloading with IRS}
%Next, we introduce a concrete example that improve the performance of task offloading latency and energy consumption. We consider the implementation of IRS. In order to further enhance the uplink offloading performance of the resource-limited UEs, great attentions have been drawn to the technology of reconfigurable intelligent surface (RIS) recently, due to its advantages of low cost, easy deployment, fine-grained passive beamforming and directional signal enhancement or interference nulling.
Next, we will introduce a concrete example of implementing IRS in multi-tier computing systems to reduce task offloading latency and energy consumption, as shown in Fig.~\ref{MMI}. Each user could either offload its task to the multi-tier nodes such as edge/fog server for computation via the IRS or to the cloud via the IRS and massive MIMO node.
In order to further improve uplink task offloading performance for resource-limited end users, IRS technology has attracted extensive attention due to its advantages of low cost, easy deploymentation, fine-grained passive beamforming, and directional signal enhancement or interference nulling. By controlling surface reflective elements, IRS can be reconfigured to provide a more favorable wireless propagation environment for communications. Obviously, using IRS in multi-tier computing systems is an economical and environmentally friendly method to facilitate task offloading \cite{2021-Wang-IRS}.

%In the next generation radio standard, the communication frequency exceeds $6$ Gigahertz. Since the high-frequency signal is sensitive to the blocking effect of trees and buildings, the coverage of the base station is significantly reduced \cite{2017-Lien-5G,2015-Bai-Mwave}. On the other hand, devices located at the cell edge or behind line of sight (LoS) blockages typically suffer from low task offload rates, which increases both the delay and energy consumption of task offloading in multi-tier computing systems \cite{2021-Bai-IRSWP}.
%IRS, regarded as effective auxiliary wireless network technology for potentially achieving high spectrum and energy efficiency via only low-cost reflecting elements, has attracted more and more attention to circumvent these restrictions, listed as one of the candidate key technologies in 6G by academia and industry \cite{2019-Wu-IRS,2020-Di-IRS,2020-Marco-RIS,2020-Wu-IRST,2021-Bai-IRSMEC,2021-Hu-IRS,2021-Wu-IRSNOMA}.
%IRS will be the key technology for next generation wireless networks, and multi-tier computing combining IRS holds great promise for many applications. In \cite{2021-Chu-IRSEd,2021-Wang-IRS}, the authors have studied the impact of the IRS on computational performance in a multi-tier computing system, which have demonstrated the benefits of the IRS to improve the computational offloading, in comparison to the benchmark schemes.
In \cite{2021-Chu-IRSEd}, Chu \MakeLowercase{\textit{et al.}} studied the impact of an IRS on computational performance in a mobile edge computing (MEC) system, targeting to optimize sum computational bits and taking into account the CPU frequency, the offloading time
allocation, transmit power of each device as well as the phase shifts of the IRS. In \cite{2021-Wang-IRS}, Wang \MakeLowercase{\textit{et al.}} investigated the task offloading problem
in a hybrid IRS and massive MIMO relay assisted fog computing system, and formulated a joint task offloading, IRS phase shift optimization, and power allocation problem to minimize the total energy consumption.
In \cite{2021-Zhou-IRS}, Zhou \MakeLowercase{\textit{et al.}} studied an IRS-assisted MEC systems, in which IRS is deployed to assist task offloading from two users to the fog/edge access point connected to the edge cloud. Under the constraint of IRS discrete phase, the authors designed the passive reflection phase of IRS and the user's computational task scheduling strategy to minimize the total task processing latency.
In \cite{2021-Bai-IRSP}, Bai \MakeLowercase{\textit{et al.}} studied an innovative framework to employ IRS in wireless powered MEC systems, and the task offloading is based on orthogonal frequency-division multiplexing (OFDM) systems. The objective is to minimize the total task offloading energy consumption. Based on the above studies, IRS is capable of providing an additional link both for data transmission and for task offloading, so as to provide enhanced computational capability.

\subsection{Task Offloading in SAGIN}
IoT seeks to connect billions of resource-constrained devices around us through heterogeneous networks. The SAGIN is viewed as a major candidate to support such IoT requirements, helping provision seamless and massive connectivity for smart services~\cite{2021-Dong-STNMEC,2021-Yu-ECSA}.
%In the past two years, the term of fifth generation (5G) networks have been commercially available and deployed across the world. Currently, 5G is still on its way, but it is already the time for academia and industry to shift their attention to 5G beyond and the sixth generation (6G) networks, in order to satisfy the future demands for intelligent IoT \cite{2021-Dong-STNMEC}. Even though discussions are ongoing as to 6G, there exists possible multi-tier computing technologies into the 6G networks from the perspectives of computing, communication and caching.
In the past two years, 5G wireless networks have been commercialized and deployed around the world. Although 5G is still in its development, academia and industry have now shifted their attention to beyond 5G and 6G wireless networks, in order to meet the demands of ultra-low latency and high energy efficiency for iIoT \cite{2021-Dong-STNMEC}. Among the discussions about 6G, from the perspective of computing, communication, and caching, it is the trend to combine SAGIN with multi-tier computing technologies in the 6G networks.

Specifically, it is widely recognized that SAGIN will be the potential core architecture of the future 6G network to support seamless and near-instantaneous hyper-connectivity \cite{2021-Dong-STNMEC}. Thus, multi-tier computing with SAGIN promotes the task offloading performance. As a key part of this, in the integrated air-ground branch, unmanned aerial vehicles (UAVs) are flexibly deployed at the aerial network layer, assisting in communication, computing and caching of ground networks due to their flexibility and proximity~\cite{2018-Cheng-AGIMEC}.
However, in 6G networks, SAGIN still faces challenges such as the demands of temporal-spatial dynamic communication/computing/caching services, large-scale complex connection decisions and resource scheduling, and ubiquitous intelligence demands within the network. To sum up, it remains extremely challenging to realize these visions of 6G in SAGIN.
%Meanwhile, it is recognized and expected that an integrated space-air-ground-underwater network will be the core potential architecture of future 6G networks to support seamless and near-instant super-connectivity \cite{2021-Dong-STNMEC}. As a critical part therein, the integrated air-ground branch, where unmanned aerial vehicles (UAVs) or drones are flexibly deployed in the air-network tier, is able to assist the communication, computing, and caching of the terrestrial network due to its flexibility and proximity \cite{2018-Cheng-AGIMEC}.
%Nevertheless, in the context of 6G networks, the air-ground integrated network is faced with several challenges, for example, temporal-spatial dynamic communication/computing/caching service demand, large scale complex connection decision and resource management, and ubiquitous intelligence demand inside the network. To summarize, it is extremely challenging to realize those visions of 6G in the air-ground integrated network.

There have been heavy research efforts on the architecture of SAGIN and multi-tier computing in the existing literature.
Cheng \MakeLowercase{\textit{et al.}} \cite{2018-Cheng-AGIMEC} proposed a novel air-ground integrated mobile edge network, by investigating the potential benefits and applications of drone cells, and UAV-assisted edge computing and caching. To support diverse vehicular services, Zhang \MakeLowercase{\textit{et al.}}~\cite{2017-Zhang-SDSAG} presented a software defined networking (SDN)-based space-air-ground integrated network architecture. Focusing on provisioning computing services by UAVs, Zhou \MakeLowercase{\textit{et al.}} \cite{2018-Zhou-AGIMEC} proposed an air-ground integrated MEC framework to cater for the urgent computing service demand from the IoTs. Furthermore, Kato \MakeLowercase{\textit{et al.}} \cite{2019-Kato-SAG} conducted a comprehensive study about how to deal with the challenges related to the space-air-ground integrated networks by AI techniques, including network control, spectrum management, energy management, routing and handover management, and security guarantee. In \cite{2019-Cheng-SAAC}, Cheng  \MakeLowercase{\textit{et al.}} demonstrated a SAGIN edge/cloud computing architecture for offloading the computation-intensive applications, considering remote energy and computation constraints, and developed a joint resource allocation and task scheduling approach to efficiently allocate the computing resources. In \cite{2020-Shang-MEC}, Shang  \MakeLowercase{\textit{et al.}} studied MEC in air-ground integrated wireless networks to minimize the total energy consumption by jointly optimizing users¡¯ association for computation offloading, uplink transmit power, allocated bandwidth, computation capacity, and UAV 3-D placement.
However, how the air network layer allocate the communication/computing/caching resources intelligently for task offloading of the ground network layer in SAGIN has not been adequately addressed.

\subsection{Edge Intelligence-Empowered Task Offloading}
With the continuing increase in the quantity and quality of rich multimedia services, the traffic and computational tasks of mobile users and smart devices have significantly increased in recent years, bringing huge workload to the already congested backbone and access networks. Even with the help of multi-tier computing systems, it is not trivial at all to satisfy the quality of experience (QoE) requirements of users. The main difficulty lies in the need of large amount of wireless data and task transmissions for task offloading, causing wireless channel congestion. Therefore, the optimization problem or decision making of the combined wireless communication resource allocation and multi-tier task offloading is the key. That is, how to share the communication resources and computing resources between edge nodes and the cloud. In response to the increasing complexity of wireless communication networks (for example, a typical 5G node is expected to have $2000$ or more configurable parameters), a recent new research trend is to optimize wireless resource allocation through AI technologies \cite{2017-Mao-MEC,2018-Mao-DL}, including but not limited to applying AI algorithms to physical layer resource allocation, data link layer resource allocation, medium access control, and traffic and congestion control \cite{2018-Tang-DL}. Especially, reinforcement learning is often applied to jointly manage communication, computing, and caching resources. With learning based multi-tier computing systems, we can optimize task offloading, communication resource allocation, and content caching at edge nodes. Further, federated learning \cite{2017-McMahan-CEDL}, as a distributed learning framework, always brings the following benefits for task offloading: 1) great reduction of the amount of data that must be uploaded through wireless uplink channel, 2) cognitive response to the changing wireless network environments and conditions, and 3) strong adaptability to the heterogeneous nodes in the wireless networks, 4) better protection of personal data privacy.

In learning-based multi-tier computing systems, task offloading decision and communication resource allocation vectors generally are binary variables, turning out it is challenging to find the optimal solution of resource allocations. Moreover, the feasible set and the objective function of the optimization problem are generally not convex, making the problem NP hard. In addition, in time-variant systems, channel conditions and computational cost are dynamic. Instead of solving the NP hard optimization problem by utilizing conventional optimization methods, the task offloading and communication resource allocation problem in multi-tier computing systems could be possibly solved using online learning algorithms. During the online learning process, the deep reinforcement learning methods might be applied to jointly optimize the subcarrier allocation and task offloading in each time episode. %In the iterative algorithm, we first solve the task scheduling subproblem by optimizing the power allocation $\mathbf{p}$, computational resource allocation $\mathbf{f}$.
Online federated learning framework is recently utilized to learn in a distributed way, in order to solve the task offloading and communication resource allocation problem. Based on the communication, computational resource allocation, the multi-tier task offloading decisions can then be optimized.

Furthermore, edge learning methods have been investigated in some edge/fog computing systems to simplify the optimization algorithm or fulfill online implementations \cite{2019-Chen-Deep,2018-Huang-Deep,2019-Wang-TVT,2020-Wang-NOMAFOG,2020-Qu-DMRO,2020-Huang-Dmec,2021-Wang-Smart,2021-Yu-DRLFL}. In \cite{2018-Huang-Deep}, Huang  \MakeLowercase{\textit{et al.}} designed a deep learning-based task offloading strategy to minimize weighted energy consumption and latency. In \cite{2021-Wang-Smart}, Wang  \MakeLowercase{\textit{et al.}} leveraged deep reinforcement learning method for smart resource allocation in a software defined network (SDN)-enabled edge computing architecture. In \cite{2020-Huang-Dmec}, Huang  \MakeLowercase{\textit{et al.}} proposed a deep learning-based task offloading strategy for offloading decisions and resource allocation of a wireless powered edge computing system. In \cite{2020-Yang-DRLIRS}, Yang \MakeLowercase{\textit{et al.}} also used deep reinforcement learning method in IRS-aided edge computing systems to enhance system security and maximize the sum rate of the down-link task offloading. In \cite{2020-Elbir-DLIRS}, a convolutional neural network was constructed for channel estimation of a large IRS-aided massive MIMO communication system to estimate the direct and the cascaded channels, used for multi-tier task offloading.

\begin{figure}[!t]
\begin{center}
\includegraphics [width=3.6in]{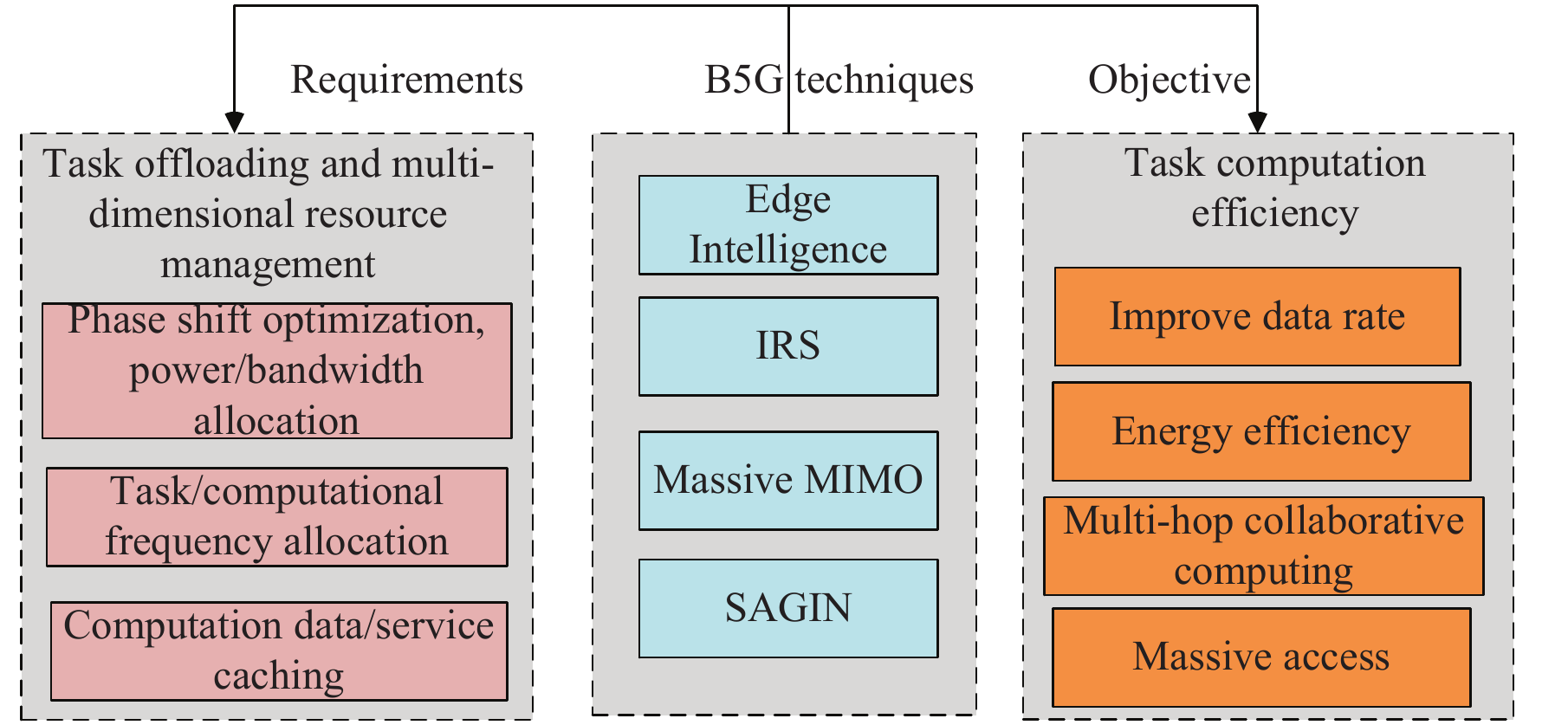}
\caption{Requirements and objective for designing multi-tier computing systems.} \label{RA}
\end{center}
\end{figure}

\section{Resource Allocation for Task Offloading}
In this section, we characterize the multi-tier computing resources allocation in next generation wireless networks. Effective optimization methods are then presented to achieve efficient task offloading with multi-tier resources.
\subsection{Main Functionality}
In this subsection, the computational and communication resources allocation, service placement, and security requirement are characterized for designing multi-tier computing systems, which is illustrated in Fig.~\ref{RA}.
\subsubsection{Computation}
Multi-tier computing architectures were envisioned to achieve rapid and affordable scalability by developing computation capabilities flexibly along the entire cloud-to-things continuum~\cite{2017-Mung-Fog}. In essence, multi-tier computing systems distribute computing capability anywhere between the cloud and the things to take full advantage of the computational resource available along this continuum, thus extending the traditional cloud computing architecture to the edge of the network. Thanks to multi-tier computing, some application components can be performed at the network's edge, like delay-sensitive components. While other components, such as time-tolerant and computation-intensive ones, are performed in the cloud. Satisfying diverse user requirements will require both cloud computing with enormous resources and distributed fog/edge computing with resources and simple algorithms closer to the users to support time-sensitive tasks. With heterogeneous computing resources and the collaborative service architecture, the proposed multi-tier computing systems are able to effectively support a full range of services in different environments. On this basis, multi-tier computing provides the advantage of low-latency task offloading since it allows task to be processed at the network edge, close to the end devices. However, cloud computing alone is not adequate for supporting all IoT applications, while a multi-tier computing system can be complementary.
%Accordingly, they help avoid resource contention at the edge of the network due to on-device data processing and cooperative radio resource management, further augmented with advanced communication, control, and storage capabilities. As a result, fog computing emerges as a unified end-to-end platform for a rapidly growing variety of dissimilar IoT services, which leverages the on-demand scalability of cloud resources as well as coordinates the involvement of geographically distributed edge and end devices. Hence, it has the potential to provide a rich set of fog computing functions for a large number of vertical IoT industries.

For smart devices with abundant computing resources, multi-tier computing seeks to achieve seamless integration of edge and cloud systems. This vision goes beyond treating the network edge and smart devices as separate computing platforms. Seamlessly integrating fleets and swarms of mobile IoT entities into a dense multi-tier enclave is a new distributed computing paradigm that improves the scalability, extensibility and assemblability of cloud services through edge of computing systems. Smart devices (cars, drones and robots) have spare computational resource, allowing the multi-tier computing platform to reduce energy consumption and task processing latency compared to the traditional edge computing scenarios relying on static and low-power edge servers.
%For intelligent IoT equipment with abundant processing and computation resources, dense fog seeks to realize their seamless integration with proximate edge equipment and remote cloud functions. This vision goes beyond treating the network edges and end devices as isolated computing platforms. Seamless integration of fleets and swarms of moving IoT entities into a dense fog enclave becomes a new distributed computing paradigm that improves scalability, extensibility, and compositionality of cloud-like services deployed closer to the network edge. In contrast to past fog-like considerations for static and low-power IoT modules (sensors, meters, actuators, and so on), more advanced capabilities of intelligent IoT devices (cars, drones, and robots) make the energy costs of collaborative fog operation truly pay off.

\subsubsection{Communication}
Multi-tier computing systems distribute communication functions anywhere between the cloud and things to take full advantage of the communication resource available along this continuum.
In massive MIMO-aided multi-tier computing systems, the achievable data rates are mostly determined by large-scale fading, and so is the communication resource allocation. This means that there is no need to frequently update communication resource allocations, hence reducing signaling overhead. IRS is capable of improving the success rate of the task offloading. Given the potential gains, if the line-of-sight (LoS) link between the task offloading nodes and computing nodes is blocked by obstacles, the task could be offloaded via the IRS reflected link. In this manner, we attempt to optimize the link selection and wireless communication resource allocation.

It is important to maintain the required data rates for task offloading. Take the task offloading from a car as an example. Given that a connected car produces tens of megabytes of data per second, an autonomous vehicle may generate up to a gigabyte per second~\cite{2016-Chiang-Fog}. Here, dense moving edge nodes can support accelerated data communication by largely utilizing directional high-rate communication in the massive MIMO, IRS or the SAGIN. Edge nodes at the same time provide novel strategies for smart devices to combine the benefits of centralized and ad-hoc topology into a unified solution by using multi-hop, multi-connection mechanisms to communicate with adjacent network infrastructure when facing the intermittent connectivity.

%Here, the dense moving fog can support the accelerated data traffic by heavily exploiting the directional high-rate communications over the mmWave bands. Dense moving fog can also provide novel ways for intelligent IoT devices to communicate with each other as well as with their proximate network infrastructure in the face of intermittent connectivity by utilizing multi-hop multi-connectivity mechanisms, thus combining the advantages of centralized and ad-hoc network topologies into a unified solution.

\subsubsection{Storage}
Because the edge nodes often have limited storage resources, distributing the data among edge and cloud nodes is vital for optimizing task offloading latency or energy consumption at a given QoS requirement. On top of it, multi-tier computing also brings a large amount of cloud-like services closer to the end users. Caching computational data or services at edge nodes is hence crucial, which relieves the burden of backhaul transmission with transporting all the data to the clouds.

Accordingly, elastic storage capacity of edge nodes might be used to support applications running on resource constrained IoT devices. Due to the inherent flexibility of multi-tier computing systems, it is possible to integrate a large number of densely distributed devices. Caching capacity of edge servers is usually accessed by both smart devices and edge access points. For example, user nodes are possibly consolidated into special capacity areas. Then, multiple interconnected edge infrastructures that coexist in space and time can pool the storage resources of adjacent edge networks together for sharing by smart devices and end users.

%Correspondingly, elastic memory capacity can be made available to various applications running on top of constrained IoT devices. Given that dimensioning of fog-aided operation is inherently flexible, the very large numbers of densely distributed and potentially mobile intelligent IoT entities may be integrated therein. Abundant storage space becomes accessible by the fog devices collectively with, for example, end nodes coalescing into ad-hoc capacity enclaves. As a result, multiple interconnected fog infrastructures that co-exist in space and time may serve as storage backup for each other by pooling various resources of the network edge, access, and end devices in proximity.

\subsubsection{Security}
In cloud computing systems, massive data needs to be uploaded to the cloud data center through a front-haul link, where data security cannot be guaranteed. However, multi-tier computing systems presents unique security challenges and opportunities. Dense edge nodes with established dynamic trust chains are acting as a trusted authority for other smart devices and systems. In particular, multi-tier computing systems with edge and cloud can handle responsibilities such as trusted computing platforms, and secure storage of short-term sensitive information. Multi-tier computing systems also use edge systems to facilitate local threat monitoring, detection, and protection for users and provide powerful proximity-based authentication services for better authentication through proxy smart devices.
%Not limited to the above angles, fog infrastructures promise unique security-related opportunities. Massive and dense moving fog with already established dynamic chains of trust can act as a trusted authority for external devices and systems. In particular, the moving fog may handle the responsibilities of a trusted computation platform, a certification authority, and a secure storage for short-lived sensitive information, among many others. Fog can also facilitate localized threat monitoring, detection, and protection for its nodes as well as offer powerful proximity-based authentication mechanisms by proxying the end devices for better identity verification.

However, the multi-tier computing systems meanwhile incur new security vulnerabilities, mainly from multiple heterogeneous nodes. For example, in a multi-node environment, when multiple potentially competing service providers and consumers share resources distributed across a set of hardware platforms, advanced authorization and authentication mechanisms need to be created to effectively leverage this heterogeneous medium and devices between edge and cloud entities. Fortunately, a trusted execution environment supported by a public key infrastructure may be a suitable solution to the above problems. Nevertheless, the intelligent integration of hardware assistance and software security mechanisms in multi-tier computing systems remains an open research challenge.
%The central concern here is that of heterogeneity: multiple potentially competing service providers and consumers are utilizing distributed and dissimilar resources across a diverse collection of hardware platforms in multi-tenant environments. Therefore, advanced authorization and authentication mechanisms need to be coined, which will effectively leverage this heterogeneous medium and mediate between the fog entities. Fortunately, trusted execution environments supported by public-key infrastructures may become a suitable remedy for the above issues. Still, intelligent integration of hardware-assisted and software-centric security mechanisms remains an open research challenge for the envisioned dense moving fog.

Additionally, multi-tier computing systems will have to cope with changing environments compared to existing edge computing systems that operate mainly under known conditions. In this case, the security mechanism for multi-tier computing systems is supposed to constantly adapt to the changing operating conditions. To address this challenge, multi-tier computing systems must dynamically adjust their overall security posture. It requires the design of new security protocols, which is able to respond to any security threat without causing service disruptions and to achieve secured and uninterrupted operation of the task offloading.
%In contrast to the state-of-the-art systems primarily operating in known conditions, the prospective dense moving fog will have to handle volatile environments. Therefore, the employed security mechanisms have to continuously adapt to the current operating conditions. Facing this challenge, dense fog has to dynamically adjust its overall security level, which calls for designing new security protocols that will be ready to respond adequately to any security compromises without creating disruptions hampering safe and uninterrupted system operation.

\subsection{Optimization Algorithms}
In this subsection, the effective optimization methods are presented to achieve efficient task offloading with multi-tier resources.
\subsubsection{Nonconvex Optimization}
During task offloading process, most of the resource allocation problems in multi-tier computing systems need to solve a series of nonconvex optimization problems. For example, for IRS-aided multi-tier computing systems, there are four blocks of optimization variables, namely, the task offloading ratio, power allocation at the relay node, and IRS phase shifts of two hops task transmission. The optimization of the task offloading ratio is related to the computing setting, while the optimization of power allocation and phase-shift matrices affects the communication design. However, the resource allocation problem in IRS-enabled multi-tier computing systems is difficult to solve due to two aspects. The first one is the coupling effect between the power allocation vector and the IRS phase-shift vector. The second one is that the objective function~(OF) is non-convex with respect to the phase shifts. Obviously, it is an open challenge to obtain a globally optimal solution directly. In fact, alternating optimization technique is a widely applicable and efficient approach for solving optimization problems involving coupled optimization variables, which has been successfully applied to several communication resource allocation problems such as hybrid precoding \cite{2016-Yu-AO}, power allocation \cite{2015-Zhao-D2D}, and IRS phase shift optimization \cite{2019-Wu-IRS,2020-Bai-MECIRS}. In this case, a locally optimal solution is usually provided. To be specific, the resource allocation optimization problem can be transformed into a phase shift optimization problem, a power allocation problem, and a task allocation problem, respectively, by using the popular alternate optimization technique to decouple communication and computational design.
%Subsequently, the optimal solutions can be provided for the power allocation $\textbf{p}$ and for task offloading ratio $\bm{\rho}$, after they are decoupled from IRS phase shifts using the alternating optimization technique. In fact, alternating optimization technique is a widely applicable and empirically efficient approach for handling optimization problems involving coupled optimization variables. It has been successfully applied to several wireless communication design problems such as hybrid precoding \cite{2016-Yu-AO}, resource allocation \cite{2015-Zhao-D2D}, and IRS-enabled wireless communication \cite{2019-Wu-IRS,2020-Bai-MECIRS}.

Remarkably, in contrast to the alternate optimization technique, distributed optimization algorithms for non-convex optimization have appeared in the literature \cite{2021-Ren-DGO}.
In \cite{2017-Tatarenko-NONCONVEX,2018-Zeng-nonconvex}, Tatarenko \MakeLowercase{\textit{et al.}} and Zeng \MakeLowercase{\textit{et al.}} studied distributed gradient descent methods for unconstrained non-convex optimization problems, respectively. %\cite{2013-Zhu-NONCONVEX,2017-Wai-NONCONVEX} propose distributed algorithms for multi-agent nonconvex problems with constraints known to all agents. \cite{2019-Farina-NONCONVEX} presents a fully asynchronous distributed approach for tackling optimization problems with local equality and inequality constraints, in which both the objective function and the local constraints can be nonconvex. In the presence of a coupled constraint, the feasible region of each agent¡¯s decision variable is influenced by other agents¡¯ decision variables.
Distributed optimization algorithms are generally divided into two categories: discrete time algorithms and continuous time algorithms. The existing work mainly focuses on discrete time algorithms, while continuous time problem has attracted extensive attention in recent years, mainly because of the wide application of continuous time setting in practical systems and the development of continuous time control technology. In addition, discrete time and continuous time algorithms are closely related to each other due to the time scale transformation.  Specifically, when the time step size approaches zero, the optimization algorithm for discrete time system is similar to the continuous time system.
Note that coupled non-linear constraints are also an important constraint in distributed optimization problems. However, distributed algorithms dealing with coupled non-linear constraints are basically convex problems, namely, both the objective function and the constraint are convex.
In \cite{2022-Wang-caching}, Wang \MakeLowercase{\textit{et al.}} studied distributed augmented Lagrangian based algorithms for non-convex optimization problems of multi-tier computing systems subject to local constraints and coupled non-linear equality constraints, and investigated the joint design of the task offloading, service caching and power allocation to minimize the total task scheduling delay.

\subsubsection{Mixed-Combinatorial Optimization}
Combinatorial optimization problems have been analyzed in many works (e.g., \cite{2009-Sergienko-combinatorial,2010-Hulyanytskyi-combinatorial,2013-Barbolina-combinatorial,2006-Yemets-LFO}). Under the framework of combinatorial optimization, an important trend is analyzing combinatorial optimization problem within the framework of Euclidean combinatorial optimization, whose optimization is carried out in a Euclidean space.
In \cite{2013-Barbolina-combinatorial,2006-Yemets-LFO}, Barbolina \MakeLowercase{\textit{et al.}} and Yemets \MakeLowercase{\textit{et al.}} studied the Euclidean combinatorial optimization problems, and investigated the properties of its convex hull and methods of solving separate classes of Euclidean problems of combinatorial optimization problems. Additionally, the general permutation set problem is an important Euclidean combinatorial optimization problem.

As previously mentioned, the resource allocation problems in multi-tier computing systems involve optimizing computation, communication and caching. In general, task offloading, task data caching and communication resource allocations are binary variables. %Thus, finding the optimal solution for resource allocation optimization problem is a combinatorial optimization problem.
Specifically, in multi-tier computing for next-generation wireless networks, we need to jointly optimize the subcarrier and bandwidth allocation \cite{2021-Chen-FL,2016-Chen-Offloading,2016-Chen-JODRA}, transmit power and receive beamforming \cite{2020-Wang-NOMAFOG,2020-Wang-MMIMO}, passive beamforming at IRS \cite{2021-Wang-IRS}, device selection \cite{2021-Wang-IRS,2018-Yang-MEETS}, location updates task offloading \cite{2019-Wang-TVT}, and computational frequency control \cite{2018-Yang-MEETS}, so as to reduce the latency and energy consumption in the task offloading procedure. Therefore, these resource allocation schemes can be formulated as a mixed combinatorial optimization problem that requires joint optimization of continuous value variables (e.g., beamforming, power control) and discrete value variables (e.g., task allocation, service placement, subcarrier allocation).
%All of these resource allocation schemes can be formulated as a mixed combinatorial optimization problem, which needs to jointly optimize continuous-valued variables (e.g., beamforming and power control) and discrete-valued variables (e.g., device selection and subcarrier allocation).

It should be noted that the existing optimization methods for mixed combinatorial optimization problems are mainly based on traditional iterative optimization approaches
\cite{2018-Ding-NOMAMEC,2019-Zheng-MCC,2019-Ren-MEC,2021-Wang-IRS,2020-Sheng-NOMAMEC},  or adopt a direct end-to-end online learning approaches \cite{2021-Wang-DRLMEC,2019-Wang-TVT}. However, they may not achieve good trade-off between algorithm complexity and resource allocation performance.
Additionally, reinforcement learning (RL)-based approaches are often involved to solve combinatorial optimization problems that are unconstrained or have few constraints due to feasibility issues  \cite{2020-Yang-NOMAMEC,2020-Wang-NOMAFOG}.
Deep RL requires a Markov process to achieve satisfactory resource allocation performance \cite{2016-Goodfellow-DL}. However, Markov process may not exist in practical combinatorial optimization problems, as they have many non-convex constraints with memory. This results in difficult design of reward features for Markov optimization process, unfeasible solutions, and potential degradation of overall performance.

\subsubsection{Stochastic Optimization}
In multi-tier computing systems, stochastic optimization approach only relies on the probabilistic description of the uncertainty of the computation capacity and radio channel condition, and is able to provide a trade-off between conservatism and probabilistic assurance for the achievable task offloading performance. Stochastic programming has been widely studied in the past decade due to its wide application in machine learning and resource allocation. In a stochastic optimization problem, the objective function or constraints are the expectation of some function of random variables (such as estimated computation capacity and channel condition in learning approach) \cite{2021-Cao-OSO}.
%Owing to its broad application in machine learning and signal processing, stochastic programming has been studied extensively over the past decade. In stochastic programs, the objective and/or constraint functions are the expectation of some function of the control variables and random variables (e.g., estimated parameters and data samples in machine learning) \cite{2021-Cao-OSO}.
The challenge of stochastic optimization is that the distribution of the random variables is often unknown. Most existing literature on stochastic programming assumes that the basic distribution of random variables is fixed and that independent samples are sequentially drawn from this common distribution. However, the basic distribution of random variables involved in stochastic optimization may change slowly over time in many practical applications. %For instance, in machine learning, the joint distributions of the input feature and label are unknown to the learner and need to be learned. Most existing works on stochastic programming presume that the underlying distribution of the random variables is fixed (i.e., does not vary with time) and independent samples are drawn from this common distribution sequentially. In contrast, in many applications, the distributions involved in the stochastic programs may vary slowly across time.
%For instance, in online learning, the unknown parameters to be estimated may change over time (e.g., tracking a moving target), which leads to time-varying joint distributions of the feature and label. Therefore, we are motivated to study stochastic optimization with time-varying distributions in this article.

Stochastic optimization in state-based systems with discrete or continuous time are often modeled with Markov chains. Their effective optimization method is an important research topic. The Markov model has a wide range of applications, especially in the area of task offloading in multi-tier computing systems.
%Their effective evaluation and optimization is an important research topic. This type of model has a wide range of applications, especially in the field of task unloading.
%State-based systems with stochastic behavior and discrete or continuous time are often modelled with the help of Markov chains. Their efficient evaluation and optimization is an important research topic. There is a wide range of application areas for such kind of models, coming especially from the field of task offloading.
Specifically, some work modeled the task offloading problem as a stochastic programming problem, and jointly optimized the task allocation and the communication resources allocation \cite{Liu}. However, in all these works, system parameters need to be acquired offline, which is impossible for a time-varying system \cite{MukherjeeM}. It should be noted that there are multiple dynamic parameters in multi-tier computing systems. Therein, user mobility and channel condition are intrinsic features of wireless networks when nodes are usually in a mobility state. Then, due to changes in network topology, these parameters are time-varying, and the stochastic task offloading framework is considered as a method of online learning where users can learn time-varying system parameters. %in a more realistic scenario, offloading decision is time varying from the changeable network topologies.
%Thus, it is critical to take the users' mobility pattern into account. Although our previous work\cite{YYT} simply proposed a basic framework and no queueing and task and resource allocation are considered.

In many multi-tier computing applications, optimization criteria are trade-offs between several competing goals, such as computational cost minimization and profit maximization. In general, in this tradeoff model, it is important to establish an optimal strategy that may often not be intuitive. However, there are also optimization problems with no tradeoff characteristics, leading to counterintuitive optimal strategies. Therefore, the use of Markov decision process (MDP) to optimize stochastic systems should not be ignored.
%In many applications, the optimization criteria are a trade-off between several competing goals, e.g. minimization of running cost and maximization of profit at the same time. For sure, in these kinds of trade-off models, it is important to establish an optimal policy which in most cases is not intuitive. However, there are also examples of target functions with no trade-off character (e.g. pure lifetime maximization) which can also lead to counterintuitive optimal policies. Therefore, using MDPs for optimization of stochastic systems should not be neglected, even if a heuristically established policy seems to be optimal.
It should be noted that there exists extensive literature in online learning task offloading for stochastic optimization, such as deep RL (DRL) \cite{2019-Wang-MDP,2019-Ismaeel-fog,2020-Xiong-MDPedge,2020-Zhang-MDP}, which generally target at a broader set of learning problems in MDPs. However, as far as we know, high-dimensional action spaces are still an urgent and challenging problem in DRL. To make the problem tractable, the general optimization problem is reduced into an MDP that only considers a meaningful parameter.
Furthermore, the Multi-Armed Bandit~(MAB) problem is a special case of MDP problems for which regret learning frameworks are generally considered to be more efficient in terms of computational complexity. Additionally, the use of the MAB model is appropriate and recognizable, taking advantage of the fact that the resources of edge node are limited. Based on the above analysis, MDP promises an online learning framework for learning computing resources and available resources information for stochastic optimization, aiming to minimize task offloading cost.
%Nonetheless, to our knowledge, the high-dimensional action space remains an urgent and challenging problem in DRL. However, Multi-Armed Bandit~(MAB) problems constitute a special class of MDPs, for which the regret learning framework is generally viewed as more effective in terms of computational complexity.
%In addition, by exploiting the fact that the edge nodes have finite resources, the use of MAB models is highly appropriate and identified. Based on the above analysis, in this paper, we propose an online learning framework to learn the computing resources and the queueing information of available fog nodes around user equipments (UEs) in dynamic fog networks, which aims to minimize the task offloading latency combining with the Combinatorial Multi-Armed Bandits~(CMAB) framework. Based on the problem formulation and theoretical analysis, we jointly optimizing the task decision allocation and the spectrum scheduling to realize the ultra-low latency requirement for delay-sensitive applications.

\section{Research Directions and Open Problems}
In this section, we present the research directions and open problems for task offloading in next generation wireless networks, supported by the wireless network infrastructures in Section II.
\subsection{Multi-Dimensional Resource Management}
Compared to cloud computing, the edge nodes and end users in multi-tier computing systems may have limited resources. Therefore, communication, computing and caching resource allocation is a very important research issue in multi-tier computing systems. Specifically, next-generation wireless communication networks present various technologies, including massive MIMO, IRS, SAGIN, and AI etc. The new computing model by combining them with multi-tier computing will reduce task offloading delay, grant large-scale user access and promote rapid development of the intelligent services, as well as enable efficient task offloading to realize efficient collaborative computing and multi-dimensional communication, caching, computation resource sharing.

However, the computation power of multi-tier servers is typically limited. The wireless physical layer resource allocation and user access techniques are the key challenges that hinder the success of multi-tier computing for 5G and beyond in executing compute-intensive and latency-critical applications. The optimization of resource allocation may be multi-objective in different situations, e.g., diverse nature of applications, heterogeneous server capabilities, user demands and characteristics, and channel connection qualities.

%Mobility: Due to dynamic nature, the EU and fog nodes leave (join) from (to) a fog layer arbitrarily. Thus, the dynamic load balancing and task assignment to mobile end device become very difficult due to the scalability issue in fog layer. Therefore, how to balance the load towards cost-effective concerning delay, power consumption, and bandwidth resource distribution is one of the open research directions. Since fog computing is localized, understanding the mobility pattern of end devices may be helpful for task assignment and resource management in fog computing.

%Virtualization: Small-scale DCs at the network edge can offload the computing task and storage requirements, after that, reduce the traffic over the network and infrastructure cost compared to large remote DCs. Although, these edge DCs benefits in many aspects such as better QoS and low communication delay, virtualization is one of the key approaches to support VDCs with different service requirements and objectives.

\subsection{Multi-Tier Task Allocation}
Since multi-tier computing systems provide extra computing capability at the network edge, one of the core problems is how to manage task allocation. More specifically, how to decide which tasks should be performed on end-user devices, at fog/edge systems, or in the cloud. At a more granular level, the challenge is to which computing nodes should a task be assigned. To achieve low latency and high energy efficiency task offloading, computing tasks need to be scheduled to computing nodes with different capabilities according to different task computing models, communication bandwidths and channel qualities. Therefore, heterogeneity becomes an important factor in multi-tier computing architectural design. Dealing with different task computation and various communication protocols to manage task offloading becomes a major problem.
%As the bottom-most layer in fog computing architecture consists of various end devices such as a smart-phone, smart-watch, virtual sensor node, intelligent devices including autonomous car, smart home devices, the heterogeneity issue arises regarding data collection, data format, and data processing capability. Nevertheless, fog node in fog cluster comprises by routers, switches, gateway and other devices with different computing and communication facilities.
%Thus heterogeneity becomes an important design factor in fog computing architecture. Handling of different data formats and various communication protocols for managing semi- or unstructured data becomes major issues.

\subsection{Heterogeneous QoS Management}
With the development of various novel technologies, intelligent services are increasingly applied in many fields of human life, including business, manufacturing, health-care, entertainment, etc. On one hand, the number of smart services deployed around edge and cloud servers is growing rapidly. On the other hand, different service providers provision services with similar functions, and different edge servers may possess different service performance. Then, the smart devices will require services with different QoS requirements. In light of these descriptions, intelligent services are migrating to the network, i.e., to edge servers residing near end users.

Note that the QoS requirements in multi-tier computing systems include task response time, throughput, reliability and availability, typically different for different users.
%Specifically, end users expect to run intelligent services with guaranteed QoS.
However, user mobility and different server capabilities turn the applicability of traditional QoS management inapplicable. Therefore, how to monitor and manage QoS attributes, and schedule multi-dimensional resources timely and effectively to fulfill specific QoS requirement for each user becomes the main issue in multi-tier computing systems.

\subsection{Data Privacy}
In multi-tier computing systems, data and computation task need to be collected close to the physically distributed edge devices, and there exists a large number of devices in the systems. When analyzing sensitive information from distributed nodes, data privacy requirements must be satisfied.
We should select computing nodes in a way that best protect the data privacy, considering the computing nodes in different parts of the network may have different privacy protection capabilities. The tasks collected, transmitted, and processed at the edge or in the cloud needs to be anonymized \cite{2017-Syed-security}. Then, multi-tier data analysis and processing is achieved securely in multi-tier computing systems. %However, in order not to affect any interconnect processes that use this information (for example, intrusion detection systems and anomaly detection algorithms), care must be taken to remove information from the data while maintaining the required level of anonymity.
Note that distributed systems are in general more vulnerable to be attacked than centralized systems, and both the devices and edge nodes in multi-tier computing systems are generally less powerful than the cloud. Therefore, these nodes may not have as adequate resources as the cloud to protect themselves. In addition, the devices and edge systems may not have enough intelligence and capability needed to detect threats due to limited resources. In all, data privacy of multi-tier computing from things to the cloud will be the focus of future research in multi-tier computing systems.
%While cloud operates in heavily protected facilities selected and controlled by cloud operators, fog often needs to operate in more vulnerable environments¡ªwhere they can best meet customer requirements and often wherever users want them to be. Many fog systems will be significantly smaller than clouds (e.g., a fog node on a vehicle, in a manufacturing plant, or on a oil rig), and hence, may not have as much resources as the clouds to protect themselves. Furthermore, each fog system may not have the global intelligence necessary for detecting threats.

\section{Conclusions}
In this paper, we investigated the key wireless communication techniques, effective resource allocation approaches and research directions to embrace the era of task offloading for multi-tier computing-based next-generation wireless networks. In particular, the multi-tier computing system model, multi-tier computing resources and optimization methods were presented for better serving the task offloading. We hope that this paper will serve as a valuable reference and guide to further promote the theoretical, algorithmic, and systematic development and advancement of the task offloading with multi-tier computing resources in next generation wireless networks.
%We hope that this article will serve as a valuable reference and guideline for further considering edge AI opportunities across theoretical, algorithmic, systematic, and entrepreneurial considerations to embrace the exciting era of edge AI.

\bibliography{mybib}
\bibliographystyle{ieeetr}

\end{document}